\documentclass[12pt]{article}

\usepackage{latexsym}

\usepackage{graphicx}

\begin{document}


\title{Charged black holes on a Kaluza-Klein bubble}

\author{
   Jutta Kunz$^{1}$\thanks{E-mail: kunz@theorie.physik.uni-oldenburg.de},  Stoytcho  Yazadjiev$^{2}$ \thanks{E-mail: yazad@phys.uni-sofia.bg}\\
{\footnotesize $^{1}$   Institut f\"ur Physik, Universit\"at Oldenburg, Postfach 2503 D-26111 Oldenburg, Germany }\\
{\footnotesize $^{2}$ Department of Theoretical Physics,
               Faculty of Physics, Sofia University,}\\
{\footnotesize  5 James Bourchier Boulevard, Sofia~1164, Bulgaria }\\
}

\date{}

\maketitle

\begin{abstract}
We construct a solution of two black holes on a Kaluza-Klein bubble
in Einstein-Maxwell-dilaton theory.
We explore the consequences of the presence of charge
for the properties of this solution,
and obtain a generalized Smarr relation and first law.
\end{abstract}


\sloppy

\section{Introduction}

In recent years the study
of the higher dimensional Einstein equations has led to
interesting new types of solutions.
Study of the generalized Weyl solutions
by Emparan and Reall \cite{Emparan:2001wk}, for instance,
has led to the construction of black ring solutions
\cite{Emparan:2001wn}
and their generalizations
(see e.g.~\cite{Emparan:2006mm} for a review).

The generalized Weyl solutions have been classified in terms
of their rod structure \cite{Emparan:2001wk}.
They also include multi black hole solutions,
analyzed in more detail by Tan and Teo \cite{Tan:2003jz},
as well as sequences of Kaluza-Klein (KK) bubbles and black holes,
analyzed by Elvang and Horowitz \cite{Elvang:2002br}
and Elvang, Harmark and Obers \cite{Elvang:2004iz}.

The presence of KK bubbles leads to interesting features
for the black hole solutions. Black hole solutions
with spherical horizon topology can, for instance, become arbitrarily
large in the presence of a bubble \cite{Elvang:2002br}.
In contrast, because of the finite size of the circle at infinity,
without a bubble such KK black holes are limited in size,
as demonstrated by Kudoh and Wiseman \cite{Kudoh:2004hs}.

The bubbles also provide the means, to hold the black holes apart,
allowing for multi black hole spacetimes without conical singularites.
Small pieces of bubble are sufficient to hold large black holes
in equilibrium, i.e., they compensate the gravitational
attraction of the black holes and at the same time
prevent them from merging \cite{Elvang:2002br}.

Recently, Kastor, Ray and Traschen \cite{Kastor:2008wd}
also addressed the role of KK bubbles for the laws of black hole
mechanics. Introducing the notion of bubble surface gravity,
they obtained interesting relations, displaying an interchange
symmetry between bubble and black hole properties.

Here we address the question as to how the properties
of such Kaluza-Klein black hole-bubble spacetimes
are affected by the presence of charge.
In particular,
we consider the 5 dimensional Einstein-Maxwell-dilaton (EMD) gravity
with arbitrary dilaton coupling parameter $\alpha$.
In the limit $\alpha \rightarrow 0$, Einstein-Maxwell theory is
recovered.
We here focus on the solution of two charged
black holes on a Kaluza-Klein bubble.

In section 2 recall the action and equations of motion.
We present the exact solution of two charged
black holes on a Kaluza-Klein bubble in section 3.
We analyze the horizon and global properties of this solution
in section 4. In section 5 we derive Smarr-like formulae
for the mass and tension and present the first law.
We address the extremal limit in section 6 and compare
with charged black strings in section 7. We present
our conclusions in section 8.

\section{EMD gravity}

We consider the 5-dimensional Einstein-Maxwell-dilaton (EMD) gravity
with action
\begin{eqnarray}
S= \int d^5x\sqrt{-g}\left(R
 - 2g^{\mu\nu}\partial_{\mu}\varphi\partial_{\nu}\varphi
 - e^{-2\alpha\varphi}F_{\mu\nu}F^{\mu\nu} \right).
\end{eqnarray}
and arbitrary dilaton coupling parameter $\alpha$.

The field equations are given by
\begin{eqnarray}\label{EMDFE}
&&R_{\mu\nu}= 2\partial_{\mu}\varphi\partial_{\nu}\varphi +2e^{-2\alpha\varphi}\left(F_{\mu\sigma}F_{\nu}^{\,\sigma}
- {1\over 6}g_{\mu\nu}F_{\lambda\sigma}F^{\lambda\sigma}\right) \nonumber \\ \nonumber \\
&&\nabla_{\mu}\left(e^{-2\alpha\varphi}F^{\mu\nu} \right)=0 , \\ \nonumber \\
&&\nabla_{\mu}\nabla^{\mu}\varphi = - {\alpha\over 2}e^{-2\alpha\varphi}F_{\sigma\lambda}F^{\sigma\lambda} .\nonumber
\end{eqnarray}

The 5D EMD gravity has a dual description in terms
of the 3-form field strength $H$ defined by
\begin{eqnarray}
H= e^{-2\alpha\varphi}\star F
\end{eqnarray}
where $\star$ denotes the Hodge dual.
In terms of the 3-form field strength $H$
the action can be written as follows
\begin{eqnarray}
S= \int d^5x\sqrt{-g}\left(R - 2g^{\mu\nu}\partial_{\mu}\varphi\partial_{\nu}\varphi
-{1\over 3} e^{2\alpha\varphi}H_{\mu\nu\sigma}H^{\mu\nu\sigma} \right).
\end{eqnarray}

The corresponding field equations of the dual theory are
\begin{eqnarray}\label{DUALFE}
&&R_{\mu\nu}= 2\partial_{\mu}\varphi\partial_{\nu}\varphi +e^{2\alpha\varphi}\left(H_{\mu\sigma\lambda}H_{\nu}^{\,\sigma\lambda} - {2\over 9}g_{\mu\nu}H_{\rho\lambda\sigma}H^{\rho\lambda\sigma}\right), \nonumber \\ \nonumber\\
&&\nabla_{\mu}\left(e^{2\alpha\varphi}H^{\mu\nu\lambda}\right)=0 , \\\nonumber \\
&& \nabla_{\mu}\nabla^{\mu}\varphi = {\alpha\over 6} e^{2\alpha\varphi}H_{\rho\sigma\lambda}H^{\rho\sigma\lambda}. \nonumber
\end{eqnarray}

\section{Exact solution}

As an explicit example of the sequences of
Kaluza-Klein (KK) bubbles and black holes \cite{Elvang:2004iz},
let us consider the solution describing
two vacuum black holes on a Kaluza-Klein bubble discussed
by Elvang and Horowitz \cite{Elvang:2002br}.
This vacuum solution has metric
\begin{eqnarray}
ds^2= -e^{2U_{t}} dt^2 + e^{2U_{\psi}}d\psi^2
     + e^{2U_{\phi}}d\phi^2 + e^{2\nu}(d\rho^2 + dz^2)
\end{eqnarray}
where
\begin{eqnarray}
&&e^{2U_{t}}=  {(R_{2} -\zeta_{2})(R_{4}-\zeta_{4})\over (R_{1} -\zeta_{1}) (R_{3} -\zeta_{3})}, \\
&&e^{2U_{\psi}}= (R_{1} -\zeta_{1})(R_{4}+\zeta_{4}),\\
&&e^{2U_{\phi}}= {R_{3}-\zeta_{3}\over R_{2} -\zeta_{2} },\\
&&e^{2\nu}= {Y_{14}Y_{23}\over 4R_{1}R_{2}R_{3}R_{4}} \left({Y_{12}Y_{34}\over Y_{13}Y_{24}} \right)^{1/2} {R_{1} -\zeta_{1}\over R_{4}-\zeta_{4} }
\end{eqnarray}
and
\begin{eqnarray}
&&R_{i}=\sqrt{\rho^2 + \zeta^2_{i} },\\
&&Y_{ij} = R_{i}R_{j} + \zeta_{i}\zeta_{j} + \rho^2,\\
&&\zeta_{1}= z-a,\,\,\, \zeta_{2}= z-b, \,\,\, \zeta_{3}= z+ b ,\,\,\, \zeta_{4}=z+c .
\end{eqnarray}

The metric functions
$U_t$, $U_\phi$ and $U_\psi$ may be regarded as Newtonian potentials
produced by line masses along the $z$-axis with densities $1/2$.
These effective sources
or rods characterize the solution \cite{Emparan:2001wk}.
The rod structure of this solution is shown in Fig.~1.
The two finite rods $-c<z<-b$ and $b<z<a$ for the $t$ coordinate
correspond to event horizons of the spacetime,
while the finite rod $-b<z<b$ for the $\phi$ coordinate
corresponds to the bubble.
The two semi-infinite rods $z<-c$ and $z>a$
for the $\psi$ coordinate then ensure, that all sources
add up to an infinite rod along the $z$-axis, as required \cite{Emparan:2001wk}.
\begin{center}
\includegraphics[width=0.6\textwidth]{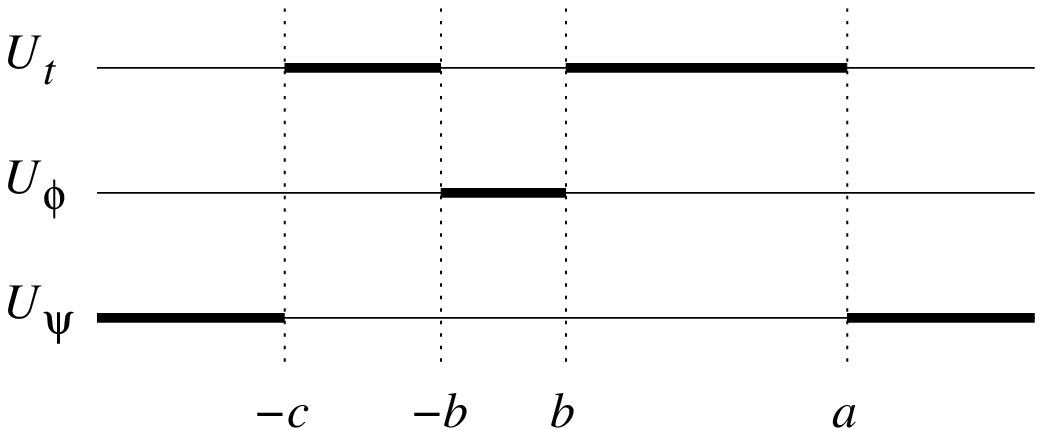}

\small{Fig.~1: The thin lines denote the $z$-axis, and the thick lines
denote the effective sources along this axis
for the metric functions $U_t$, $U_\phi$ and $U_\psi$.}
\end{center}

Using the method of \cite{Yazadjiev:2005hr}
we generate a new EMD solution from the above vacuum solution
\cite{Elvang:2002br}.
This solution then describes two charged $S^3$-black holes
sitting on a Kaluza-Klein bubble in the EMD gravity.
For $\alpha=0$ it reduces to a solution of
Einstein-Maxwell gravity.
The spacetime metric of the solution is given by the line element
\begin{eqnarray}
&&ds^2 = -{e^{2U_{t}}\over [\cosh^2\gamma -e^{2U_{t}}\sinh^2\gamma]^{2\over 1+\alpha^2_{5}} }dt^2 \nonumber \\
&& +
[\cosh^2\gamma -e^{2U_{t}}\sinh^2\gamma]^{1\over 1+\alpha^2_{5}} \left[e^{2U_{\psi}}d\psi^2 + e^{2U_{\phi}}d\phi^2 + e^{2\nu}(d\rho^2
+ dz^2) \right]
\end{eqnarray}
The electric potential $\Phi$ and the dilaton field are given by
\begin{eqnarray}
\Phi = {\sqrt{3}\tanh\gamma\over 2\sqrt{1 + \alpha^2_{5}}} {1- e^{2U_{t}}\over 1 - \tanh^2\gamma e^{2U_{t}} },\\
e^{-\alpha\varphi}= \left[\cosh^2\gamma - e^{2U_{t}}\sinh^2\gamma \right]^{\alpha_{5}^2\over 1 + \alpha^2_{5} }
\end{eqnarray}
where $\alpha_{5}= {\sqrt{3}\over 2}\alpha$
and $\gamma$ is an arbitrary parameter. The Maxwell 2-form is respectively

\begin{eqnarray}
F=-d\Phi \wedge dt.
\end{eqnarray}

\section{Analysis of the solution }

\subsection{ Asymptotic behavior }

In order to study the asymptotic behaviour of the solution
we introduce the asymptotic coordinate $r$  defined by
\begin{eqnarray}
r=\sqrt{\rho^2 +z^2}.
\end{eqnarray}
Then in the asymptotic limit $r\to \infty$ we find
\begin{eqnarray}
&&g_{tt} \approx -1 +{a +c -2b\over r}\left(1 + 2{\sinh^2\gamma\over 1 + \alpha^2_{5}} \right),\\
&&g_{\phi\phi}\approx 1  + {1\over r}\left[(a+c-2b){\sinh^2\gamma\over 1 + \alpha^2_{5}} -2b\right],\\
&&g_{\psi\psi}\approx \rho^2 \left(1 - {1\over r}\left[(a+c-2b){\sinh^2\gamma\over 1 + \alpha^2_{5}} -(a+c)\right] \right) ,\\
&&\Phi \approx {\sqrt{3}\over 2}{\cosh\gamma\sinh\gamma\over \sqrt{1+ \alpha^2_{5}}} {(a+c-2b)\over r} ,\\
&&e^{-\alpha\varphi} \approx 1 + {\alpha^2_{5} \sinh^2\gamma\over 1 + \alpha^2_{5}} {(a+c-2b)\over r}.
\end{eqnarray}

The leading asymptotic of the metric is

\begin{eqnarray}
ds^2\sim -dt^2 + \rho^2d\psi^2 + d\rho^2 + dz^2 + d\phi^2.
\end{eqnarray}
In other words, asymptotically  the space-time is ${\cal M}^4\times S^1$. The compact dimension is parameterized by $\phi$. In what follows
we will consider spacetimes with a fixed length $L$ of the Kauza-Klein circle at infinity.

\subsection{Regularity}

The rod structure of the solution in general entails conical
singularities. To cure such conical singularities at the location
of the rods, the associated coordinate must be periodic
with a particular period.

* For the semi-infinite rods $(-\infty,c]$ and $[a,+\infty)$
corresponding to the $\psi\psi$ part of the metric
regularity requires the period
\begin{eqnarray}
\Delta \psi= 2\pi\lim_{\rho\to 0} \sqrt{\rho^2g_{\rho\rho}\over g_{\psi\psi}}= 2\pi
\end{eqnarray}

* For the finite rod $[-b,b]$ corresponding to the $\phi\phi$ part of the metric
regularity requires the period
\begin{eqnarray}
\Delta \phi= 2\pi\lim_{\rho\to 0} \sqrt{\rho^2g_{\rho\rho}\over g_{\phi\phi}}= 8\pi {b(a+c)\over\sqrt{(a+b)(b+c)} }=L,
\end{eqnarray}
where $L$ is the length of the KK circle at infinity.

\subsection{Horizons}

\subsubsection{First black hole horizon}

The first black hole horizon is located at $\rho=0$ for $-c< z<-b$.
The metric of the spatial cross section of the horizon
is given by
\begin{eqnarray}
ds^2_{h}= \cosh^{2\over 1 + \alpha^2_{5}}\gamma \left[ {z+b\over z-b }d\phi^2 + 4(a-z)(z+c)d\psi^2 \right.\nonumber \\ \left. + { (z-b)\over (z+b)(a-z)(z+c) }
{(a+c)^2 (c-b)\over (c+b) } dz^2\right]. \nonumber
\end{eqnarray}
The orbits of $\phi$ and $\psi$ shrink to zero size at $z=-b$ and $z=-c$, respectively. This shows that the black hole horizon
topology is $S^3$.

By the direct calculation we find  the horizon area
\begin{eqnarray}\label{firstbherea}
{\cal A}^{(1)}_{h}&=& 32\pi^2 \cosh^{3\over 1 + \alpha^2_5}\gamma  {b(a+c)^2 (c-b)\over (b+c) }\sqrt{c-b\over a+b} \nonumber \\
&=&4\pi  \cosh^{3\over 1 + \alpha^2_5}\gamma  L (a+c)(c-b)\sqrt{c-b\over c+b }
\end{eqnarray}
The temperature of the horizon can be obtained from the surface gravity
and is given by
\begin{eqnarray}
T^{(1)}= {1\over 4\pi}{\cosh^{-3\over 1+ \alpha^2_5}\gamma   \over a + c } \sqrt{c+b\over c-b}=
\cosh^{-3\over 1+ \alpha^2_5}\gamma {2b\over L\sqrt{(a+b)(c-b)}}.
\end{eqnarray}

\subsubsection{First black hole near horizon metric}

The near horizon geometry of the left black hole
can be illustrated by performing the
coordinate transformation \cite{Tan:2003jz}
\begin{eqnarray}
\displaystyle
\rho&=& {1\over 2} \sqrt{ 1- {2(c-b)\over R^2} } \ R^2 \sin 2 \theta, \\
z &=& -{1\over 2} \left( 1- {c-b\over R^2} \right)  \ R^2 \cos  2 \theta
\ . \end{eqnarray}
The near horizon metric then becomes
\begin{eqnarray}
ds^2 &=& - \cosh^{-4\over 1 + \alpha^2_{5}}\gamma \,
{f_b(\theta)\over f_a(\theta)}
 \left( 1- {2(c-b)\over R^2} \right) dt^2 \nonumber \\
&& + \cosh^{2\over 1 + \alpha^2_{5}}\gamma \,
 {f_b(\theta)\over f_a(\theta)} \left\{
{2 (a+c)^2\over b+c} \left[ \left( 1- \frac{2(c-b)}{R^2} \right)^{-1}
dR^2 + R^2 d\theta^2 \right] \right. \nonumber \\
&&   \hspace{3.5cm}
\left. + {f_a(\theta)\over 2 f_b^2(\theta)} R^2 \cos^2 \theta d \phi^2
 + {2 f_a^2(\theta)\over f_b(\theta)} R^2 \sin^2 \theta d \psi^2 \right\}
\end{eqnarray}
with
\begin{eqnarray}
f_b(\theta) &=& 2b+(c-b) \cos^2 \theta,  \nonumber \\
f_a(\theta) &=& a+b +(c-b) \cos^2 \theta \ ,
\end{eqnarray}
corresponding to a distorted black hole.

\subsubsection{Second black hole horizon}

The second black hole horizon is located at $\rho=0$ for $b< z< a$.
The metric of its spatial cross section is
\begin{eqnarray}
ds^2_h =  \cosh^{2\over 1 + \alpha^2_{5}}\gamma \left[ {z-b\over z+b }d\phi^2 + 4(a-z)(z+c)d\psi^2  \right. \nonumber \\ \left.
+{(z+b)\over (z-b)(a-z)(z+c) }  {(a+c)^2(a-b)\over (a+b) }dz^2 \right].
\end{eqnarray}
The orbits of $\phi$ and $\psi$ shrink to zero size at $z=b$ and $z=a$, respectively showing that the
horizon topology is $S^3$.

The  horizon area is
\begin{eqnarray}\label{secondbharea}
{\cal A}^{(2)}_{(h)}&=& 32\pi^2 \cosh^{3\over 1 + \alpha^2_5}\gamma  {b(a+c)^2 (a-b)\over  (a+b)}\sqrt{a-b\over b+c}  \nonumber \\
&=& 4\pi \cosh^{3\over 1 + \alpha^2_5}\gamma  L (a+c)(a-b)\sqrt{a-b\over a+b } .\nonumber
\end{eqnarray}
and its temperature
 \begin{eqnarray}
T^{(2)}= {1\over 4\pi}{\cosh^{-3\over 1+ \alpha^2_5}\gamma   \over a + c } \sqrt{a+b\over a-b}=
\cosh^{-3\over 1+ \alpha^2_5}\gamma {2b\over L \sqrt{(b+c)(a-b)} } .
\end{eqnarray}

The near horizon metric for the second black hole can be found from the near horizon metric of the first black hole by interchanging
the parameters $a$ and $c$. The near horizon metric shows that the second black holes is distorted by the interaction with the first black
holes and the Kaluza-Klein bubble.

 \subsubsection{Bubble}

The bubble is located at  $\rho=0$ for $-b< z< b$.
The metric on the bubble for $t=const$ and $\rho=0$ is given by
\begin{eqnarray}
ds^2_{bubble}= \left[\cosh^2\gamma - {(b^2-z^2)\over (a-z)(z+c)}\sinh^2\gamma \right]^{1\over 1+ \alpha^2_5}
\left[4(a-z)(z+c)d\psi^2  \right.\nonumber  \\ \\ \left.+ {4b^2 (a+c)^2 \over (a+b)(b+c) } {dz^2\over b^2-z^2 }\right].
\nonumber
\end{eqnarray}
As it is clear form the rod structure diagram and as one sees from the bubble metric, the orbits of $\psi$ do not vanish at
$z=\pm b$. So the bubble has cylindrical topology.

 \subsubsection{Distance between black holes}

The proper distance along a curve with $\psi=const$  between the black holes
is given by
\begin{eqnarray}\label{distance}
l_{12}= {2b(a+c)\cosh^{1\over 1+\alpha^2_5}\gamma\over \sqrt{(a+b)(b+c)} }\int^{b}_{-b}  \left[1 - {(b^2-z^2)\over (a-z)(z+c)}\tanh^2\gamma \right]^{1\over 2(1+ \alpha^2_5)} {dz\over \sqrt{b^2-z^2}} .
\end{eqnarray}
It can be shown that
\begin{eqnarray}
{1\over 4}L \le l_{12}\le {1\over 4}L \cosh^{1\over 1+\alpha^2_5}\gamma .
\end{eqnarray}

\subsection{Masses}

\subsubsection{ADM mass}

In order to compute the ADM mass of the solution we use the general
results of \cite{Deser:1988fc} and \cite{Townsend:2001rg}.
We then find
\begin{eqnarray}
M_{ADM} &=& {1\over 2}L\left[a+c-b
 + {3\sinh^2\gamma\over 2(1+\alpha^2_{5})}(a+c-2b) \right]  \nonumber \\
 &=& 4\pi {b(a+c)\over \sqrt{(a+b)(b+c)} }\left[a+c-b
 + {3\sinh^2\gamma\over 2(1+\alpha^2_{5})}(a+c-2b) \right]
\label{mass}
\end{eqnarray}

With the asymptotic form of the relevant metric components given by
\begin{eqnarray}
\label{1}
g_{tt}\simeq -1+\frac{c_t}{r},
~~~g_{\phi\phi}\simeq 1+\frac{c_\phi}{r}
\label{asym}
\end{eqnarray}
the mass (\ref{mass}) can be directly read off from
the asymptotics of the metric potentials
\cite{Harmark:2003dg,Kol:2003if,Harmark:2004ch}
\begin{eqnarray}
\label{2}
M_{ADM}=\frac{1}{4}L (2c_t-c_\phi).
\end{eqnarray}

The ADM mass can also be calculated via the generalized Komar integral
introduced in \cite{Townsend:2001rg}
\begin{eqnarray}\label{Komar}
M^{K}=-{L\over 16\pi} \oint_{\infty} dS^{\mu\nu\sigma}\left(2\eta_{\sigma}\nabla_{\mu}\xi_{\nu} +\xi_{\nu}\nabla_{\mu}\eta_{\sigma} \right)
\end{eqnarray}
where $dS^{\mu\nu\sigma}$ is the surface element
and $\xi={\partial/\partial t}$,  $\eta={\partial/\partial \phi}$ are the time translation and the Killing vector associated with the fifth dimension,
respectively. Here the integration is performed over a 2-dimensional
sphere at spatial infinity of ${\cal M}^4$.

\subsubsection{Intrinsic masses}

When dealing with multi-object configurations it is useful
to properly define the intrinsic mass of
each object in the configuration. As in four dimensions, the natural definition of the intrinsic mass of each black hole is the Komar
integral evaluated on the horizon of the corresponding black hole
\begin{eqnarray}
M_{i}^{{\cal H}}=-{L\over 16\pi} \oint_{{\cal H}_i} dS^{\mu\nu\sigma}\left(2\eta_{\sigma}\nabla_{\mu}\xi_{\nu} +\xi_{\nu}\nabla_{\mu}\eta_{\sigma} \right).
\end{eqnarray}
Here the integration is over a 2-dimensional surface which is an intersection of the horizon with a constant $t$ and $\phi$ hypersurface.

The intrinsic masses of the black holes for the solution under consideration are given by
\begin{eqnarray}
M^{{\cal H}}_{1}= {1\over 2}L(c-b)=4\pi {b(a+c)(c-b)\over \sqrt{(a+b)(b+c)}}={1\over 4\pi}\kappa_{1} {\cal A}^{(1)}_{h}
={1\over 2}T^{(1)}_{h} {\cal A}^{(1)}_{h},
\end{eqnarray}

\begin{eqnarray}
M^{{\cal H}}_{2}={1\over 2}L(a-b) =4\pi {b(a+c)(a-b)\over \sqrt{(a+b)(b+c)}}={1\over 4\pi}\kappa_{2} {\cal A}^{(2)}_{h}
={1\over 2}T^{(2)}_{h} {\cal A}^{(2)}_{h},
\end{eqnarray}
where $\kappa_{i}$ is the surface gravity of the $i$-th black hole.

\subsubsection{Bubble mass}

In the same way we can define the intrinsic mass
of the Kaluza-Klein bubble
\begin{eqnarray}
M_{bubble}= -{L\over 16\pi} \oint_{bubble} dS^{\mu\nu\sigma}\left(2\eta_{\sigma}\nabla_{\mu}\xi_{\nu} +\xi_{\nu}\nabla_{\mu}\eta_{\sigma} \right)
=
4\pi {b^2(a+c)\over\sqrt{(a+b)(b+c)} }={1\over 2}Lb.
\end{eqnarray}
The integration is over the 2-dimensional boundary of a constant $t$ and $\phi$ hypersurface at the bubble.

It is  worth noting that the intrinsic mass of the bubble is given by
\begin{eqnarray}
M_{bubble}={L\over 8\pi} \kappa_{b}{\cal A}^{(b)}
\end{eqnarray}
where $\kappa_{b}$ is the so-called bubble surface gravity
and ${\cal A}^{(b)}$ is the bubble area.
These quantities have been introduced in \cite{Kastor:2008wd}
and will be discussed in more detail in the next section.
In our case the bubble is smooth, i.e.,
there are no conical singularities at the bubble.
For such bubbles one obtains \cite{Kastor:2008wd}
\begin{eqnarray}
\kappa_{b}={2\pi\over L},
\end{eqnarray}
and
\begin{eqnarray}
{\cal A}^{(b)}= 2Lb,
\end{eqnarray}
where  $L$ is the length of the Kaluza-Klein circle at infinity.
Note, that $\kappa_{b}$ and ${\cal A}^{(b)}$ are not affected
by the presence of charge.

\subsection{Electric charge}

The electric charge of the $i$-th black hole is defined by
\begin{eqnarray}
Q_{i}= {1\over 2\pi} \oint_{{\cal H}_{i}} e^{-2\alpha\varphi}\star F= {1\over 2\pi}  \oint_{{\cal H}_{i}} H.
\end{eqnarray}
Explicit calculations give
\begin{eqnarray}
Q_{i}= {\sqrt{3}\cosh\gamma\sinh\gamma\over \sqrt{1+ \alpha^2_{5}}} M^{\cal H}_{i}.
\end{eqnarray}
Thus the individual black holes have the same charge to mass ratio.

Denoting by $\Phi^{{\cal H}}_{i}$ the electric potential evaluated
on the horizon ${\cal H}_{i}$, we obtain explicitly
\begin{eqnarray}
\Phi^{{\cal H}}_{1}=\Phi^{{\cal H}}_{2}={\sqrt{3}\tanh\gamma\over 2\sqrt{1 +\alpha^2_5}}.
\label{PhiH}
\end{eqnarray}

\subsection{Dilaton  charge}

Finally, we define the dilaton charge $\Sigma$ via the asymptotic
behaviour of the dilaton field
\begin{equation}
\varphi \approx   \frac{\Sigma}{r}
 = - \frac{\alpha^2_5 \sinh^2 \gamma \, (a+c -2b)}{\alpha ( 1+ \alpha^2_5)\, r } .
\end{equation}
Comparison with the electric charges and potential then yields the
relation
\begin{equation}
Q_{1}\Phi^{{\cal H}}_{1} + Q_{2}\Phi^{{\cal H}}_{2}
= - L \frac{\Sigma}{\alpha} .
\label{Sigma}
\end{equation}
The dilaton charge thus represents no independent parameter.

\section{Smarr-like relations and first law}

\subsection{Mass formula}

It can be checked that our  solution satisfies
the following Smarr-like relation
\begin{eqnarray}\label{Smarr}
M_{ADM}= M^{{\cal H}}_{1} +   Q_{1}\Phi^{{\cal H}}_{1} + M^{{\cal H}}_{2} +   Q_{2}\Phi^{{\cal H}}_{2} + M_{bubble} .
\end{eqnarray}
Here Eq.~(\ref{Sigma}) can be used to replace the electromagnetic
contributions by the dilaton contribution.

Below we give a general proof of the Smarr-like relation (\ref{Smarr}).
Our starting point is the generalized Komar integral (\ref{Komar}).
Using the Gauss theorem  we find that the ADM-mass can be written as
a sum of a bulk integral over a constant $t$ and $\phi$ hypersurface and surface integrals on the black holes horizons
and the bubble
\begin{eqnarray}
&&M_{ADM}=-{L\over 16\pi} \oint_{\infty} dS^{\mu\nu\sigma}\left(2\eta_{\sigma}\nabla_{\mu}\xi_{\nu} +\xi_{\nu}\nabla_{\mu}\eta_{\sigma} \right)= \nonumber\\ \nonumber\\
&&{L\over 8\pi}\int_{\Sigma} dS^{\mu\nu}\left[2\eta_{\nu}R_{\mu\lambda}\xi^{\lambda} -\xi_{\nu}R_{\mu\lambda} \eta^{\lambda}\right]
-\sum_{i}{L\over 16\pi} \oint_{{\cal H}_i} dS^{\mu\nu\sigma}\left(2\eta_{\sigma}\nabla_{\mu}\xi_{\nu} +\xi_{\nu}\nabla_{\mu}\eta_{\sigma}\right)
\nonumber \\ \\
&&
-{L\over 16\pi} \oint_{bubble} dS^{\mu\nu\sigma}\left(2\eta_{\sigma}\nabla_{\mu}\xi_{\nu} +\xi_{\nu}\nabla_{\mu}\eta_{\sigma}\right). \nonumber
\end{eqnarray}

On the horizons we have
\begin{eqnarray}\label{HSE}
dS_{\mu\nu\sigma}= {3!\over \sqrt{g(\eta,\eta)}} \xi_{[\mu}N_{\nu}\eta_{\sigma]} d\omega,
\end{eqnarray}
where $d\omega$ is the magnitude of the 2-dimensional surface element and $N_{\nu}$
is the second null normal to the horizon
normalized such as $\xi^{\nu}N_{\nu}=-1$.
Substituting into the horizon surface integrals and taking into account that $\xi$ and $\eta$ are commuting  we obtain
\begin{eqnarray}
-{L\over 16\pi} \oint_{{\cal H}_i} dS^{\mu\nu\sigma}\left(\xi_{\nu}\nabla_{\mu}\eta_{\sigma}\right)=0
\end{eqnarray}
and therefore
\begin{eqnarray}
&&-{L\over 16\pi} \oint_{{\cal H}_i} dS^{\mu\nu\sigma}\left(2\eta_{\sigma}\nabla_{\mu}\xi_{\nu} +\xi_{\nu}\nabla_{\mu}\eta_{\sigma}\right)=
-{L\over 16\pi} \oint_{{\cal H}_i} dS^{\mu\nu\sigma}\left(2\eta_{\sigma}\nabla_{\mu}\xi_{\nu}\right)=\nonumber \\\\
&&-{L\over 4\pi} \oint_{{\cal H}_i}d\omega  \sqrt{g(\eta,\eta)} N_{\nu}\xi_{\mu}\nabla^{\mu}\xi^{\nu}.
\end{eqnarray}
By definition, on the horizons
$\xi_{\mu}\nabla^{\mu}\xi^{\nu}=\kappa_{i}\xi^{\nu}$,
where $\kappa_{i}$ is the surface gravity of the $i$-th horizon.
Hence we obtain
$N_{\nu}\xi_{\mu}\nabla^{\mu}\xi^{\nu}=\kappa_{i}N_{\nu}\xi^{\nu}
= - \kappa_{i}$
and therefore
\begin{eqnarray}
&&-{L\over 16\pi} \oint_{{\cal H}_i} dS^{\mu\nu\sigma}\left(2\eta_{\sigma}\nabla_{\mu}\xi_{\nu} +\xi_{\nu}\nabla_{\mu}\eta_{\sigma}\right)=
-{L\over 4\pi} \oint_{{\cal H}_i}d\omega  \sqrt{g(\eta,\eta)} N_{\nu}\xi_{\mu}\nabla^{\mu}\xi^{\nu} = \nonumber \\ \\
&&{\kappa_{i}L\over 4\pi} \oint_{{\cal H}_i}d\omega \sqrt{g(\eta,\eta)}
\end{eqnarray}
where we have made use of the fact that the surface gravity is constant
on the horizon.

Further, taking into account that
\begin{eqnarray}
L \oint_{{\cal H}_i}d\omega \sqrt{g(\eta,\eta)}= \oint_{{\cal H}_i}d\omega d\phi \sqrt{g(\eta,\eta)}={\cal A}_{h}^{(i)}
\end{eqnarray}
is the area of the $i$-th horizon we find that
\begin{eqnarray}
-{L\over 16\pi} \oint_{{\cal H}_i} dS^{\mu\nu\sigma}\left(2\eta_{\sigma}\nabla_{\mu}\xi_{\nu} +\xi_{\nu}\nabla_{\mu}\eta_{\sigma}\right)=
{1\over 4\pi} \kappa_{i}{\cal A}_{h}^{(i)}= {1\over 2}T^{(i)}{\cal A}_{h}^{(i)}=M^{{\cal H}}_{i}.
\end{eqnarray}

Let us denote by $n^{1}_{\mu}$ and $n^{2}_{\mu}$
the unit spacelike normals to the bubble surface.
The bubble surface gravity is defined
by the following equality on the bubble (see \cite{Kastor:2008wd})
\begin{eqnarray}
\nabla_{[\mu}\eta_{\nu]}=\kappa_{b}(n^{1}_{\mu}n^2_{\nu}-n^{1}_{\nu}n^2_{\mu} ).
\end{eqnarray}
As shown in \cite{Kastor:2008wd},
the bubble surface gravity $\kappa_{b}$ is constant on the bubble surface. In our case this is obvious since  $\kappa_{b}=2\pi/L$.
For the bubble surface element we can write
\begin{eqnarray}
dS_{\mu\nu\sigma}= {3!\over \sqrt{|g(\xi,\xi)|}}\xi_{[\mu}n^1_{\nu}n^2_{\sigma]}d\omega.
\end{eqnarray}
Substituting in the bubble surface integral and taking into account that $\eta$ vanishes on the bubble we have
\begin{eqnarray}
 \oint_{bubble} dS^{\mu\nu\sigma}\left(2\eta_{\sigma}\nabla_{\mu}\xi_{\nu}\right)=0
\end{eqnarray}
and therefore we find
\begin{eqnarray}
&&M_{bubble}=-{L\over 16\pi} \oint_{bubble} dS^{\mu\nu\sigma}\left(2\eta_{\sigma}\nabla_{\mu}\xi_{\nu}
+\xi_{\nu}\nabla_{\mu}\eta_{\sigma}\right)= -{L\over 16\pi} \oint_{bubble} dS^{\mu\nu\sigma}\left(\xi_{\nu}\nabla_{\mu}\eta_{\sigma}\right)= \nonumber \\ \nonumber\\ &&{L\kappa_{b}\over 8\pi}  \oint_{bubble}\sqrt{|g(\xi,\xi)|}d\omega= {L\over 8\pi}\kappa_{b}{\cal A}^{(b)}
\end{eqnarray}
where
\begin{eqnarray}
{\cal A}^{(b)}= \oint_{bubble}\sqrt{|g(\xi,\xi)|}d\omega
\end{eqnarray}
is the bubble area as defined in \cite{Kastor:2008wd}.

Let us now evaluate the bulk integral
\begin{eqnarray}
I_{bulk}= {L\over 8\pi}\int_{\Sigma} dS^{\mu\nu}\left[2\eta_{\nu}R_{\mu\lambda}\xi^{\lambda} -\xi_{\nu}R_{\mu\lambda} \eta^{\lambda}\right].
\end{eqnarray}
Using the field equations (\ref{EMDFE}) we obtain
\begin{eqnarray}
I_{bulk}= {L\over 2\pi}\int_{\Sigma} dS^{\mu\nu}\left[e^{-2\alpha\varphi}\left(F_{\mu\sigma}\xi^{\lambda}F_{\lambda}^{\,\sigma}- {1\over 4}F^2\xi_{\mu} \right)\eta_{\nu} -{1\over 2}e^{-2\alpha\varphi} F_{\mu\sigma}\eta^{\lambda}F_{\lambda}^{\,\sigma}\xi_{\nu} \right].
\end{eqnarray}

Making use of the Lie symmetries ${\cal L}_{\xi}A=0$ and ${\cal L}_{\eta}A=0$ as well as the field equations one can show that the following relations are satisfied
\begin{eqnarray}
&&e^{-2\alpha\varphi} F^{\mu}_{\,\,\sigma}\xi^{\lambda}F_{\lambda}^{\,\sigma}= - \nabla_{\sigma}\left(e^{-2\alpha\varphi}F^{\mu\sigma}\xi^{\lambda}A_{\lambda} \right), \nonumber \\\nonumber \\
&&e^{-2\alpha\varphi} F^{\mu}_{\,\,\sigma}\eta^{\lambda}F_{\lambda}^{\,\sigma}= - \nabla_{\sigma}\left(e^{-2\alpha\varphi}F^{\mu\sigma}\eta^{\lambda}A_{\lambda} \right), \\\nonumber \\
&&{1\over 4}e^{-2\alpha\varphi}F^2\xi^{\mu}= \nabla_{\sigma}\left(e^{-2\alpha\varphi} \xi^{[\mu}F^{\sigma]\lambda}A_{\lambda} \right). \nonumber
\end{eqnarray}
Using these relations and the fact that the integrand has vanishing Lie derivative with respect to the Killing vector $\eta$ the bulk integral can be presented as a sum of surface integrals by applying the Gauss theorem
\begin{eqnarray}\label{SURFINT}
I_{bulk}= \sum_{i} {L\over 4\pi} \oint_{{\cal H}_i} dS_{\mu\nu\sigma}\left[e^{-2\alpha\varphi}\left(F^{\mu\sigma}\xi^{\lambda}A_{\lambda} + \xi^{\mu}F^{\sigma\lambda}A_{\lambda} \right)\eta^{\nu} - {1\over 2} e^{-2\alpha\varphi}F^{\mu\sigma}\eta^{\lambda}A_{\lambda}\xi^{\nu} \right]
\nonumber \\ \nonumber\\
+ {L\over 4\pi} \oint_{bubble} dS_{\mu\nu\sigma}\left[e^{-2\alpha\varphi}\left(F^{\mu\sigma}\xi^{\lambda}A_{\lambda}
+ \xi^{\mu}F^{\sigma\lambda}A_{\lambda} \right)\eta^{\nu} - {1\over 2} e^{-2\alpha\varphi}F^{\mu\sigma}\eta^{\lambda}A_{\lambda}\xi^{\nu} \right]
 \\\nonumber \\
-  {L\over 4\pi} \oint_{\infty} dS_{\mu\nu\sigma}\left[e^{-2\alpha\varphi}\left(F^{\mu\sigma}\xi^{\lambda}A_{\lambda}
+ \xi^{\mu}F^{\sigma\lambda}A_{\lambda} \right)\eta^{\nu} - {1\over 2} e^{-2\alpha\varphi}F^{\mu\sigma}\eta^{\lambda}A_{\lambda}\xi^{\nu} \right].
\nonumber
\end{eqnarray}

The Killing vector $\eta$ vanishes on the bubble and therefore the surface integral on the bubble vanishes too. The same holds for the
surface integral at infinity since we consider a gauge in which  $A_{\lambda}$ vanishes at infinity. As a consequence we find that only the first term in
(\ref{SURFINT}) survives. Further we consider pure electric solutions satisfying
\begin{eqnarray}
F^{[\mu\sigma}\xi^{\nu]}=0 , \; \; \eta^{\mu}F_{\mu\nu}=0.
\end{eqnarray}
Under these conditions and using the 2-dimensional horizon surface element (\ref{HSE}) we find that
\begin{eqnarray}
&&I_{bulk}= \sum_{i}{L\over 4\pi} \oint_{{\cal H}_i} (-\xi^{\lambda}A_{\lambda}) e^{-2\alpha\varphi} F^{\mu\sigma}
(\xi_{\mu}N_{\sigma}- \xi_{\sigma}N_{\mu} )\sqrt{g(\eta,\eta)}d\omega = \nonumber \\ \nonumber  \\
&&\sum_{i}{1\over 4\pi} \oint_{{\cal H}_i} (-\xi^{\lambda}A_{\lambda}) e^{-2\alpha\varphi} F^{\mu\sigma}
(\xi_{\mu}N_{\sigma}- \xi_{\sigma}N_{\mu} )\sqrt{g(\eta,\eta)}d\phi d\omega = \\ \nonumber  \\
&&\sum_{i}{1\over 4\pi} \oint_{{\cal H}_i} (-\xi^{\lambda}A_{\lambda}) e^{-2\alpha\varphi} F^{\mu\sigma} dS_{\mu\sigma}
 \nonumber
\end{eqnarray}
where we have taken into account that
\begin{eqnarray}
(\xi_{\mu}N_{\sigma}- \xi_{\sigma}N_{\mu} )g(\eta,\eta)d\phi d\omega = (\xi_{\mu}N_{\sigma}- \xi_{\sigma}N_{\mu} )d{\cal A}= dS_{\mu\nu}
\end{eqnarray}
is the 3-dimensional surface element on the horizon.

In the final step we use the fact that the potential $\Phi=-\xi^{\lambda}A_{\lambda}$ is constant\footnote{The fact that the potential
$\Phi$ is constant on the horizon can be shown by the well-known manner. On the horizon we have $R_{\mu\nu}\xi^\mu\xi^\nu=0$ which,
in view of the field equations, implies that $\partial_{\mu}\Phi$ is null on the horizon. Since $\xi_\mu$ is null on the horizon and orthogonal to
$\partial_{\mu}\Phi$ it follows that on the horizon  $\partial_{\mu}\Phi$ is proportional to $\xi_\mu$, $\partial_{\mu}\Phi=\sigma \xi_{\mu}$.
For an arbitrary vector $\zeta^\mu$ tangential to the horizon we have ${\cal L}_{\zeta}\Phi= \sigma \zeta^{\mu}\xi_{\mu}=0$ (on the horizon) which shows that $\Phi$ is
constant on the horizon.  } on the horizon and therefore we
obtain
\begin{eqnarray}
 I_{bulk}= \sum_{i}\Phi^{{\cal H}}_{i}{1\over 4\pi} \oint_{{\cal H}_i} e^{-2\alpha\varphi} F^{\mu\sigma} dS_{\mu\sigma}=
 \sum_{i}\Phi^{{\cal H}}_{i}Q_{i}.
\end{eqnarray}
In this way we have  proven the Smarr-like formula
\begin{eqnarray}
M_{ADM}=\sum_{i}M^{{\cal H}}_{i} + \sum_{i}\Phi^{{\cal H}}_{i}Q_{i} + M_{bubble}.
\end{eqnarray}

Obviously the above relation can be generalized easily for the case when more than one bubble is present in the configuration.

\subsection{Tension}

A Smarr-like  relation can also be found  for the tension. The tension can be calculated by the following Komar integral \cite{Townsend:2001rg}
\begin{eqnarray}
{\cal T} L = -{1\over 16\pi } \oint_{\infty} dS_{\mu\nu\sigma} \left(\eta^{\sigma} \nabla^{\mu}\xi^{\nu} + 2\xi^{\nu}\nabla^{\mu}\eta_{\sigma}\right).
\end{eqnarray}
With the asymptotic form of the relevant metric components given by
Eq.~(\ref{asym}), the tension can be directly read off from
the asymptotics of the metric potentials
\cite{Harmark:2003dg,Kol:2003if,Harmark:2004ch}
\begin{eqnarray}
\label{3}
{\cal T} L =\frac{1}{4}L ( c_t-2c_\phi) .
\end{eqnarray}

Following the same method as for the derivation of the Smarr-like relation for ADM mass we find
\begin{eqnarray}
{\cal T} L = {1\over 2}\sum_{i} M^{{\cal H}}_{i} + 2 M_{bubble}.
\end{eqnarray}

Combining this relation with the relation for ADM mass one can easily show that the following relation holds
\begin{eqnarray}
{\cal T} L = {1\over 2}M_{ADM} -{1\over 2}\sum_i\Phi^{{\cal H}}_iQ_i + {3\over 2}M_{bubble}. \nonumber
\end{eqnarray}

\subsection{First law}

In the absence of bubbles the first law can be derived following the well known method. In the presence of bubbles we have to
find the contribution of the bubbles to the first law. It was however shown in \cite{Kastor:2008wd}  that the smooth bubbles
(i.e. without conical singularities ) do not give any contribution to the first law. In this paper we consider namely smooth bubbles.
Therefore, in our case the first law conserves its usual form
\begin{eqnarray}\label{FirstLaw}
\delta M_{ADM}= {1\over 4}T^{(1)}\delta A^{(1)}_{h} + \Phi_{1}\delta Q_{1}
+ {1\over 4}T^{(2)}\delta A^{(2)}_{h} + \Phi_{2}\delta Q_{2} .
\end{eqnarray}
Let us note again that we keep $L$ fixed.

\section{Extremal limit}

The extremal solutions are obtained in the limit $\gamma \to \infty$, keeping the charges $Q_1$ and $Q_2$ fixed.
The explicit form of the extremal solutions  is the following
\begin{eqnarray}
ds^2&=& - \left( 1 + {q_2 \over R_2} + {q_1\over R_3}  \right)^{-{2\over 1+\alpha^2_5}}dt^2 \nonumber \\
&& + \,
\left( 1 + {q_2 \over R_2} + {q_1\over R_3}  \right)^{1\over 1 + \alpha^2_5}
 \left[(R_2-\zeta_2)(R_3+\zeta_3)d\psi^2
+ {R_3-\zeta_3\over R_2-\zeta_2} d\phi^2 \right. \nonumber \\
&& \left. + \,
{Y_{23}\over 2R_2R_3 } {R_2-\zeta_2\over R_3-\zeta_3}(d\rho^2 + dz^2) \right]
,  \\ 
\Phi &=& {\sqrt{3}\over 2 \sqrt{1+\alpha^2_5}} {{q_2\over R_2} + {q_1\over R_3}\over 1 + {q_2\over R_2} + {q_1\over R_3}  }, \nonumber \\\nonumber  \\
e^{-\alpha\varphi} &=& \left(1 + {q_2 \over R_2} + {q_1\over R_3}\right)^{\alpha^2_5\over 1+ \alpha^2_5} \nonumber
\end{eqnarray}
where the parameters $q_1$ and $q_2$ are given by
\begin{eqnarray}
q_i= 2\sqrt{1 +\alpha^2_5\over 3}{Q_{i}\over L} .
\end{eqnarray}

Strictly speaking, the extremal solutions are singular and they can not be interpreted as extremal black holes. As
one can see the dilaton field is divergent on the candidate-horizons
$R_2=0$ and $R_3=0$ for $\alpha\ne 0$.
As we shall show below, only in the Einstein-Maxwell case ($\alpha=0$) the solutions describe two extremal black holes sitting on a Kaluza--Klein bubble.

The extremal solutions can be casted in new coordinates $r$ and $\vartheta$ by performing the transformation
\begin{eqnarray}
\rho=\sqrt{r^2-2rb}\sin\vartheta, \;\;\; z=(r-b)\cos\vartheta
\end{eqnarray}
which gives
\begin{eqnarray}
ds^2&=& - \left[ 1 + {q_{2}\over r-b-b\cos\vartheta} + {q_{1}\over r-b+b\cos\vartheta}\right]^{-{2\over 1+ \alpha^2_5}} dt^2 + \nonumber \\  \nonumber\\
&& \hspace{-1cm}
\left[ 1 + {q_{2}\over r-b-b\cos\vartheta} + {q_{1}\over r-b+b\cos\vartheta}\right]^{{1\over 1+ \alpha^2_5}}
\left[{dr^2\over 1-{2b\over r} } + r^2d\vartheta^2 + r^2\sin^2\vartheta d\psi^2+ (1 - {2b\over r})d\phi^2  \right], \nonumber \\  \nonumber\\
\Phi &=& {\sqrt{3}\over 2 \sqrt{1+\alpha^2_5}}\left[ 1 - {1\over 1 + {q_{2}\over r-b-b\cos\vartheta} + {q_{1}\over r-b+b\cos\vartheta}}\right],  \nonumber \\ \\
e^{-\alpha\varphi}&=&\left[1 + {q_{2}\over r-b-b\cos\vartheta} + {q_{1}\over r-b+b\cos\vartheta} \right]^{\alpha^2_5\over 1+ \alpha^2_5}. \nonumber
\end{eqnarray}

The new coordinates $r$ and $\vartheta$ appear to be more suitable
from a physical point of view and the extremal solutions look more tractable.
In the particular case $q_2=q_1=0$ we obtain the 5D solution
describing a static KK bubble in standard spherical coordinates.
The extremal solutions can be interpreted physically as solutions
describing two charged objects
whose gravitational attraction is balanced by the electric repulsion
in the presence of a Kaluza-Klein bubble.

Let us now show that, in the Einstein-Maxwell case,
the extremal solutions describe two extremal black holes
on a Kaluza-Klein bubble.
For this purpose we consider the metric in the vicinity
of the candidate-horizon $r=2b, \vartheta=0$ and put
\begin{eqnarray}
r=2b + y^2_1 ,\;\;\;\; \vartheta=y_2
\end{eqnarray}
where $y_1$ and $y_2$ are small.
Expanding the metric around $r=2b$ and $\vartheta=0$
and performing the coordinate transformation
\begin{eqnarray}
dt=d\tau -\sqrt{2bq^3_2} d \left({1\over R}\right)
\end{eqnarray}
where $ R=y^2_1 + {b\over 2}y^2_2$ we obtain
\begin{eqnarray}\label{extrxpansion}
ds^2= -{R^2\over q^2_2} d\tau^2 - \sqrt{2b\over q_2}dRd\tau + 8bq_2 \,d\Omega^2_3.
\end{eqnarray}
Here $d\Omega^2_3$ is the metric on the unit three dimensional sphere, i.e.,
\begin{eqnarray}
d\Omega^2_3 =d\theta^2 + \sin^2\theta d\psi^2 + \cos^2\theta d\phi^2_1
\end{eqnarray}
where $\phi_1=\phi/4b$ and $\theta=\arctan(\sqrt{{b\over 2}}{y_2\over y_1})$.
Since $L=8\pi b$ in the extremal limit
the angle $\phi_1$ has the canonical period $\Delta \phi_1=2\pi$.

It is seen from (\ref{extrxpansion})
that the metric remains regular as $r\to 2b$ and $\vartheta\to 0$,
and the same is true for the electromagnetic field.
Therefore $r=2b, \vartheta=0$ is a (degenerate) event horizon
with $S^3$ topology as follows from (\ref{extrxpansion}).
The radius of the horizon is $\sqrt{8bq_2}$
and hence the area of the horizon is
\begin{eqnarray}
{\cal A}^{(2)E}_{h}(\alpha=0)= 2\sqrt{\pi} \left({ 2Q_2 \over \sqrt{3}} \right)^{3/2}.
\end{eqnarray}
The area is also found from (\ref{secondbharea}) by taking the extremal limit.

In  the same manner one can show that $r=2b,\vartheta=\pi$
is a (degenerate) event horizon with $S^3$ topology and area
\begin{eqnarray}
{\cal A}^{(1)E}_{h}(\alpha=0)= 2\sqrt{\pi} \left({ 2Q_1 \over \sqrt{3}} \right)^{3/2}.
\end{eqnarray}
The extremal limit of (\ref{firstbherea}) gives the same expression
for the erea of the first extremal black hole.
For the two extremal Einstein-Maxwell black holes sitting on a Kaluza-Klein bubble we have

\begin{eqnarray}
&&M^{E}_{bubble} =4\pi b^2={L^2\over 16\pi} , \nonumber \\ \\
&&M^{E}_{ADM}={L^2\over 16\pi}   + {\sqrt{3}\over 2}(Q_1 + Q_2) .\nonumber
\end{eqnarray}

When $\alpha\ne 0$ we consider near extremal (NE) solutions.
We here find that the area of the NE black holes tends to zero
\begin{eqnarray}
{\cal A}^{(1)NE}_{h}(\alpha\ne 0)\to 0, \;\; {\cal A}^{(2)NE}_{h}(\alpha\ne 0)\to 0.
\end{eqnarray}

The temperature of the NE black holes is
\begin{eqnarray}
&&T^{(i)}(\alpha^2_5<2)\to 0, \nonumber \\ \nonumber\\
&&T^{(i)}(\alpha^2_5=2)\to {1\over 2\sqrt{2\pi Q_i}} ,  \\ \nonumber\\
&&T^{(i)}(\alpha^2_5>2) \to \infty .\nonumber
\end{eqnarray}

The mass of the  bubble and the ADM mass are, respectively,
\begin{eqnarray}
&&M^{NE}_{bubble} \approx4\pi b^2={L^2\over 16\pi} , \nonumber \\ \\
&&M^{NE}_{ADM}\approx{L^2\over 16\pi}   + {\sqrt{3}\over 2\sqrt{1 + \alpha^2_5}}(Q_1 + Q_2) .\nonumber
\end{eqnarray}

The proper distance $l_{12}$ between the  black holes becomes arbitrary large when extremality is approached.

\section{Black strings}

For the vacuum solution the question was addressed,
whether the total area of the two black holes ${\cal A}_{2BH}$
is larger or smaller than the area of a black string ${\cal A}^s_{h}$
with the same mass \cite{Elvang:2002br}.
It turned out, that
$${\cal A}_{2BH} < {\cal A}^s_{h}$$
showing that the black string is entropically favoured.
Here, we address this question for charged black holes.

In order to generate the solution describing
a 5-dimensional black string in Einstein-Maxwell-dilaton gravity
we consider the 4-dimensional Schwarzschild solution trivially embedded in five dimensions
\begin{eqnarray}\label{emebeddedSch}
ds^2_{Sch} = -(1 -{2m\over r})dt^2 +  {dr^2 \over 1-{2m\over r} } + r^2d\Omega^2 + dx^2_{5}
\end{eqnarray}

Using the method of hep-th/0507097 and (\ref{emebeddedSch}) as a seed solution  one obtains the following  EMD solution
\begin{eqnarray}\label{blackstring}
&&ds^2 = -{1 -{2m\over r} \over  \left[1+ {2m\over r}\sinh^2\gamma_1\right]^{2\over 1 + \alpha^2_5}}dt^2 +
\left[1+ {2m\over r}\sinh^2\gamma_1\right]^{1\over 1 + \alpha^2_5}\left[ {dr^2 \over 1-{2m\over r} } + r^2d\Omega^2 + dx^2_{5}\right], \nonumber \\ \nonumber \\
&&e^{-2\alpha\varphi} = \left[1+ {2m\over r}\sinh^2\gamma_1\right]^{2\alpha^2_5\over 1 + \alpha^2_5}, \\ \nonumber \\
&&\Phi= {\sqrt{3}\over 2} {\cosh\gamma_1 \sinh\gamma_1 \over \sqrt{1+ \alpha^2_5}} {2m\over r + 2m\sinh^2\gamma_1}\nonumber
\end{eqnarray}
An alternative derivation of this solution has been
given in \cite{Kleihaus:2006ee}.

The ADM mass, the charge, the horizon area and the temperature of the
EMD black string are given by
\begin{eqnarray}
&&M^{s}_{ADM}= mL \left[1 + {3\over 2(1+ \alpha^2_5) } \sinh^2\gamma_1 \right] , \nonumber \\\nonumber \\
&&Q^{s} = {\sqrt{3}\cosh\gamma_1\sinh\gamma_1 \over \sqrt{1 + \alpha^2_5}} mL , \nonumber\\ \\
&& {\cal A}^{s}_{h}= 16\pi \cosh^{3\over 1 + \alpha^2_5}\gamma_1\, m^2 L ,\nonumber\\ \nonumber \\
&& T^{s} = {\cosh^{-3\over 1 + \alpha^2_5}\gamma_1\, \over 8\pi m} \nonumber
\end{eqnarray}
and the following Smarr-like relation is satisfied
\begin{eqnarray}
M^{s}_{ADM} = {1\over 2} {\cal A}^{s}_{h} T^{s} + \Phi_{h}Q^{s}
\end{eqnarray}

The extremal limit of the black string solution is obtained for $\gamma_1\to \infty$ and $m\to 0$ keeping the charge $Q^s$ finite. The extremal solution  is singular with zero horizon area for all coupling parameters $\alpha_5$.
The temperature of the near extremal solution tends to zero for $\alpha^2_5<1/2$, to infinity for
$\alpha^2_5>1/2$ and the temperature is finite for $\alpha^2_5=1/2$.

Comparing now the area of two charged black holes ${\cal A}_{2BH}$ with the area of a charged string
of the same mass, charge and the length $L$ of the Kaluza-Klein circle,  we find
\begin{equation}
{\cal A}_{2BH}< {\cal A}^s_{h} .
\end{equation}
This shows that the solution is globally unstable as all solutions of this type.

\section{Conclusion}

We have constructed a solution of two black holes on a Kaluza-Klein bubble
in Einstein-Maxwell-dilaton theory,
and studied the consequences of the presence of charge
for the properties of this solution.
Furthermore, we have obtained a generalized Smarr relation and first law
for this solution, extending recent vacuum results \cite{Kastor:2008wd}.

While we have focussed here on a particular solution
\cite{Elvang:2002br},
the calculations are easily extended to general
sequences of smooth Kaluza-Klein (KK) bubbles and black holes
\cite{Elvang:2004iz}.
For these we expect the Smarr relation
$$ M= \sum_i M^{{\cal H}}_i  + \sum_i \Phi^{{\cal H}}_i Q_i + \sum_i M^{i}_{bubble}   $$
and first law
$$ \delta M = {1\over 4}\sum_i T^i\delta {\cal A}^i_{h} + \sum_i \Phi^{{\cal H}}_i \delta Q_i . $$

As for all solutions of the considered
type the question of their classical stability remains open.

Future work may involve the inclusion of rotation
\cite{Tomizawa:2007mz}.
Considering the solutions from the point of view of string theory,
on the other hand, may necessitate the inclusion of
further fields for consistency.

\section*{Acknowledgements}

S.Y. gratefully acknowledges support by the Alexander von Humboldt
foundation.

\end{document}